\documentclass[aps,prl,twocolumn,nopacs,superscriptaddress,longbibliography]{revtex4-1}

\usepackage[breaklinks=true,colorlinks,citecolor=blue,linkcolor=blue,urlcolor=blue]{hyperref}
\usepackage{epsfig,mathrsfs,color,latexsym,subfigure,marginnote,graphicx,verbatim,relsize,mathrsfs,color,array,amsmath,amsfonts,amssymb,graphicx}
\begin{document}

\title{High-$T_c$ superconductors as a New Playground for High-order Van Hove singularities and Flat-band Physics}

\author{Robert S. Markiewicz}
\email{r.markiewicz@northeastern.edu}
\affiliation{Department of Physics, Northeastern University, Boston, Massachusetts 02115, USA}

\author{Bahadur Singh}
\email{bahadur.singh@tifr.res.in}
\affiliation{Department of Condensed Matter Physics and Materials Science, Tata Institute of Fundamental Research, Colaba, Mumbai 400005, India}

\author{Christopher Lane}
\affiliation{Theoretical Division, Los Alamos National Laboratory, Los Alamos, New Mexico 87545, USA}
\affiliation{Center for Integrated Nanotechnologies, Los Alamos National Laboratory, Los Alamos, New Mexico 87545, USA}

\author{Arun Bansil}
\affiliation{Department of Physics, Northeastern University, Boston, Massachusetts 02115, USA}

\begin{abstract}

Beyond the two-dimensional (2D) saddle-point Van Hove singularities (VHSs) with logarithmic divergences in the density of states (DOS), recent studies have identified higher-order VHSs with faster-than-logarithmic divergences that can amplify electron correlation effects. Here we show that the cuprate high-Tc superconductors harbor high-order VHSs in their electronic spectra and unveil a new correlation that the cuprates with high-order VHSs display higher T$_c$’s. Our analysis indicates that the normal and higher-order VHSs can provide a straightforward new marker for identifying propensity of a material toward the occurrence of correlated phases such as excitonic insulators and supermetals. Our study opens up a new materials playground for exploring the interplay between high-order VHSs, superconducting transition temperatures and electron correlation effects in the cuprates and related high-Tc superconductors.
\end{abstract} 

\maketitle

{\it Introduction.$-$} Many properties of materials are driven by their one-particle electronic density of states (DOS). A large DOS at the Fermi level implies that many electrons contribute to the low-energy phenomena so that many-body interactions are enhanced. In particular, extrema and saddle-points in band dispersions induce Van Hove singularities (VHSs) in the DOS \cite{VHS} and the associated logarithmic divergences in two-dimensions (2D) have been a focus of interest for many years. Recent studies, however, show that 2D and 3D VHSs can be anomalously strong in some cases to yield higher-order VHSs with power-law divergences \cite{Liang, Liang3, MBMB, Marsig} that can further amplify many-body interaction and drive exotic correlated phenomena such as supermetals \cite{Liang2}. Such high-order VHSs have been reported in Moire heterostructures, slow-graphene, magic-angle twisted bilayer and trilayer graphene, Sr$_3$Ru$_2$O$_7$, and other systems that host flat bands and reduced bandwidths \cite{Liang,Tgraphene,Ref1b,Ref2a,Ref2b,SRO327}. Beyond classifying the DOS anomalies of higher-order VHSs, it is important to understand their complex effects to fully appreciate their general promise of driving correlated physics in materials. 

The role of normal VHSs in the cuprates and the related high-$T_c$ superconductors has been debated for decades. One of the earliest theories of cuprate superconductivity is that it is driven by a large DOS associated with the VHSs \cite{HirSc}. However, the connections between the cuprate superconductivity, doping, and VHSs are complex and material dependent. Thus, while in the lanthanum-based cuprates, the VHS is at the Fermi level near optimal doping, in many cuprates the VHS crosses the Fermi level well into the overdoped regime. 

Over the past 35 years, evidence for what is now known as high-order VHSs has arisen in cuprates in a number of contexts, starting with an experimental observation\cite{Ref1c} in YBa$_2$Cu$_3$O$_7$ (YBCO) of a high-order VHS in what may be an unusual surface state wherein the surface remains overdoped and nonsuperconducting independent of the bulk doping\cite{Damasc1}. Further observations include Andersen's extended VHS\cite{Andex} with power-law diverence $p_V=-\partial N/\partial E$ =0.25\cite{Zehyer} (where $N(E)$ is the DOS) and a high-order VHS which may be associated with pseudogap collapse \cite{Michon}, consistent with an earlier prediction \cite{RM70}. We hope that this systematic analysis of the many ways high-order VHSs affect the cuprate phase diagram will provide guidance for their study in other materials, while shedding light on some of the many puzzles remaining in cuprates.  

Here, we explore the existence of high-order VHSs in the cuprates and discuss how such VHSs can drive competing orders and other complex effects in materials. We find that VHS singularities can exist not only in the DOS or the $Q=(0,0)$ susceptibility but also in the susceptibility at finite  momenta (e.g. $Q\sim(\pi,\pi)$). These two VHSs compete with each other and undergo independent evolutions when the material is doped or its dispersion is tuned. By tuning the dispersions, one can generate flat bands with high-order VHSs that lead to frustration rather than instabilities. We also show the existence of high-order VHSs in bosonic bands and discuss that if the susceptibility is considered as a dispersion of electron-hole pair bosons, the resulting high-order bosonic VHSs would allow a straightforward identification of excitonic phases in materials.  Importantly, we demonstrate a correlation between the superconducting transition temperature $T_c$ and the strength of the associated VHS.

Beyond the cuprates, our study has important implications in several areas of high current interest as follows.  (1) We show how the definition of higher-order VHSs can be extended to reveal its connection with the broad area of {\it flat-band} physics. (2) We demonstrate a connection between higher-order VHSs and the electronic dimensionality. (3) We show that the connection with superconductivity is indirect, implying that additional factors are involved, most likely associated with competing instabilities.

\begin{figure}[h!]
\centering
\includegraphics[width=0.99\columnwidth]{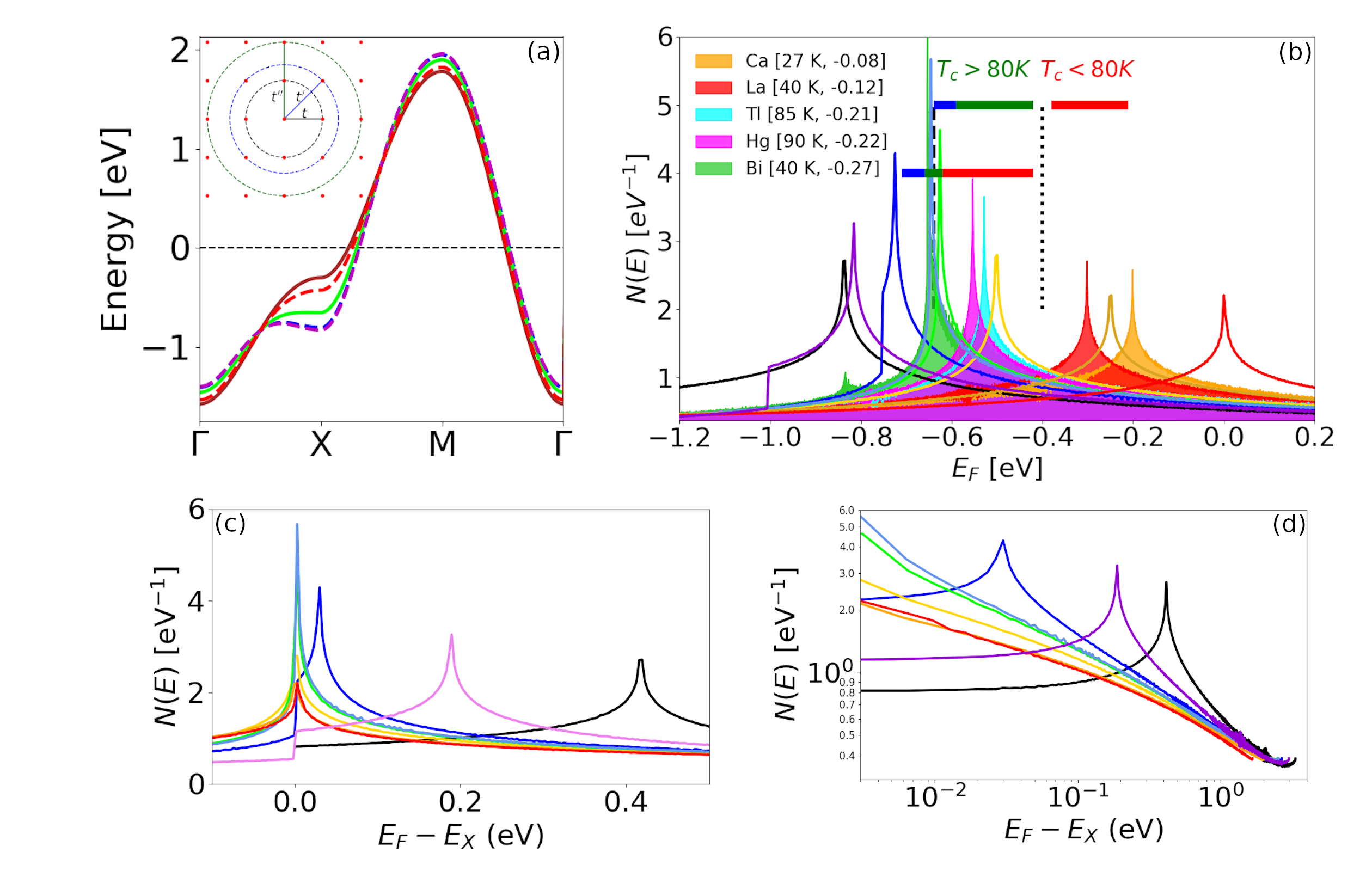}
\caption{{\bf higher order VHSs in the cuprates ($t''=-t'/2$).} (a) Dispersions for $t'$ = -0.12 (dark red), -0.17 (red dashed), -0.258 (light green -- the hoVHS), -0.32 (violet dashed), and -0.33 (blue dashed).  Inset: Definition of the hopping parameters $t$, $t'$, and $t''$.  (b) DOS $N(E)$ for several values of $t'$.   As VHS moves from right to left, the white-background curves correspond to $t'/t$ = 0 (red curve), -0.1 (orange), -0.2 (yellow-green), -0.25 (green), -0.258 (light blue), -0.3 (blue), -0.4 (violet),  and -0.5 (black), while the colored-background curves correspond to the monolayer cuprates as indicated in the legends with $t'$ values from SM S-I.  Horizontal bars indicate range of VHS peak positions for 10 bilayer or trilayer cuprates\cite{PavOK}, sorted by optimal $T_c$s, with red ($70K\ge T_c\ge 50K$), green ($100K\ge T_c\ge 90K$), and blue ($135K\ge T_c\ge 125K$) colors. The antibonding bands are above and the bonding bands below. The length of the bars indicates the range of energies over which the corresponding VHS peaks would occur.  A clear correlation of high-order VHSs with higher superconducting $T_c$s is seen.  (c,d) White-background data from frame (b) reploted as $E_F-E_X$, on linear (c) or logarithmic (d) scales.  } 
\label{fig:1vhs} 
\end{figure}

{\it High-order VHSs$-$} We begin by recalling that the energy dispersion in the cuprates can be approximated by a $t-t'-t''$ model \cite{PavOK,MBMB}

\begin{equation}
E=-2t(c_{x}+c_{y})-4t'c_{x}c_{y}-2t''(c_{2x}+c_{2y}),
\end{equation}
where $c_{nr} = cos(nk_ra)$, $a$ is the lattice constant, $r = \{x, y\}$, and the hopping parameters are defined in the inset to Fig.~\ref{fig:1vhs}(a).  In this model, $t$ sets the energy scale and thus, the evolution of energy dispersions depends on the two parameters, $t'/t$ and $t''/t$ which constitute the material dependence. We examine this energy dispersion considering $t'/t$ and $t''/t$ values relevant to cuprates and delineate high-order VHSs. Our choices of $t'$ values for different cuprates is discussed in the Supplementary Material (SM) Section S-I.  We consider an average DFT value of $t=-0.5$ eV for all the calculations.  It should be noted that nearly localized $d$- and $f$-electrons may be sensitive to just a few hopping parameters so that similar calculations should determine the characteristic properties of high-order VHSs in many correlated materials beyond cuprates.

Figure \ref{fig:1vhs}(b) presents the evolution of VHS lineshapes with $t'$ for the special value $t''=-t'/2$ that best describes most families of cuprates \cite{PavOK,MBMB}.  The corresponding VHS peak is considered as a marker of $t'$ which locates various cuprates along the $x$-axis. Such a correlation of $t'$ with the VHS allows us to compare the strength of the VHS divergence with $T_c$ for several families of cuprates. The vertical black dashed and dotted lines in Fig.~\ref{fig:1vhs}(b) delineate the range
over which cuprates with $T_c> 80K$ are found, while cuprates to the right of the dotted line have $T_c \le 80K$.  Strikingly, the high-$T_c$ cuprates
are all clustered to one side of the hoVHS, with $T_c$ increasing as the hoVHS is approached, with one exception.  The Bi-cuprates are the only cuprate family in which $T_c$ changes significantly with number of CuO$_2$-layers per unit cell, even though all have similar $t'$-values. The present results suggest a possible explanation for this anomaly.  Thus Bi2201, with maximum $T_c$ of 40K, has the highest magnitude of $t'$, placing it very close to the hoVHS -- or perhaps beyond it, given some uncertainty in $t'$.  The low $T_c$ could then follow from $|t'|$ being too large.  In Bi2212, bilayer splitting causes the effective $t'$ values to split (SM S-I), pushing the antibonding band into the range for its $T_c=$90K.  

Note further that for bilayer cuprates, the correlation holds for the antibonding band that is closer to the Fermi level, whereas the bonding bands all have $t'$s that correspond to super VHSs.  These results clearly suggest that high-order VHSs play a significant role in the superconductivity of the cuprates.
 
To further understand the evolution of VHSs, it is convenient to measure energy $E$ from the energy of the $X =(\pi,0)$ point, $i.e.$, $E_X=4 (t'-t'')$. As seen in Fig.~\ref{fig:1vhs}(c), there is always a VHS at $E_X$, which evolves from logarithmic (saddle-point) at small $t'$ to a step at larger $t'$, where the associated single Fermi surface changes into a region of Fermi surface with three pockets. The step is the point at which two pockets first appear for a given $t'$.   The crossover occurs at a critical value $t'_c$ where a pocket forms with the strongest VHS divergence (for that $t''/t'$). It has a step on the low-energy side and a power-law divergence on the high energy side.  This evolution is further illustrated by replotting the data for $E>E_X$ on logarithmic scales in Fig.~\ref{fig:1vhs}(d).  There are two types of behavior, separated by the turquoise line.  For small $|t'|$ (red to turquoise curve), the divergent peak stays at $E_X$, evolving from logarithmic to power law.  The turquoise curve has the largest, pure power-law divergence.  For larger $|t'|$, all curves (black to turquoise) start off with the same power law growth at large $\delta E=E_f-E_X$, but as $\delta E$ decreases, the curves gradually split off on realizing a power-law to  logarithmic crossover at an energy away from $E_X$. Finally, we note from Fig.~\ref{fig:1vhs}(a) that the $t'$ of strongest VHS can be approximately determined from the dispersion, as corresponding to the flattest band near $(\pi,0)$.  Such a strong VHS corresponds to Andersen's extended VHS \cite{Andex}.  The critical value of $t'$ has a simple analytic interpretation, of maximizing the degree to which the fermi surface at the VHS is tangent to the $x$- and $y$- axes -- i.e., maximizing the one-dimensionality.  Thus, for the VHS at $(0,\pi)$, $\partial E/\partial (ak_x) = 2s_x(t+2t'c_y+4t''c_x)\rightarrow 2s_x(t+2t'-4t'')$, where $s_x=sin(k_xa)$.  This vanishes when 
\begin{equation}
t'_c/t = - 1/(2-4t''/t')=-1/4,
\label{eq:2}
\end{equation}
close to the Bi cuprates.

{\it VHS Dichotomy $-$} However, our problem is far from solved.  The instabilities associated with the VHSs form a Lie group which is SO(8) for cuprates.\cite{Mikefest} For our purpose, the most important subgroup is whether the instability involves intra-VHS coupling that corresponds to $q=0$ ({\it i.e.}, a peak in the DOS) or inter-VHS coupling that yields a peak in the $Q=(\pi,\pi)$ susceptibility. These two unstable modes compete such that in the original Hubbard model the $(\pi,\pi)$ instability at half-filling is a (well-known) high-order VHS with a $ln^2$ instability which dominates the $ln$ DOS. Only with finite doping does the $(\pi,\pi)$ peak weaken as the DOS peak increases and begins to play an important role.  Thus, focusing solely on the DOS would miss the strong antiferromagnetism of cuprates. The existence of this $ln^2$ effect has been questioned since the Hubbard model requires extreme fine-tuning with all hopping parameters set to zero except nearest-neighbor $t$. However, in SM S-IV, we display a large family of dispersions with $ln^2$ susceptibility divergence.

Remarkably, the $(\pi,\pi)$ VHS is completely insensitive to the Fermi surface nesting that produces structure in the DOS, only gradually crossing over from $ln^2$ to $ln$ as the dispersion is tuned away from the Hubbard limit by either doping or tuning hopping parameters.  Hence, in general, at some hopping $t'_{cross}$ the DOS instability will become dominant.  This is accompanied by a doping $x_{cross}$ where the dominant near-$(\pi,\pi)$ instability crosses over to a near-$\Gamma$ instability (SM S-II.B). We find that optimal superconductivity falls close to $t'_{cross}$ (SM Fig.~S2) where superconductivity can tilt the balance between the two competing phases, and suggest similar behavior at $x_{cross}$.

\begin{figure}[h!]
\centering
\includegraphics[width=0.99\columnwidth]{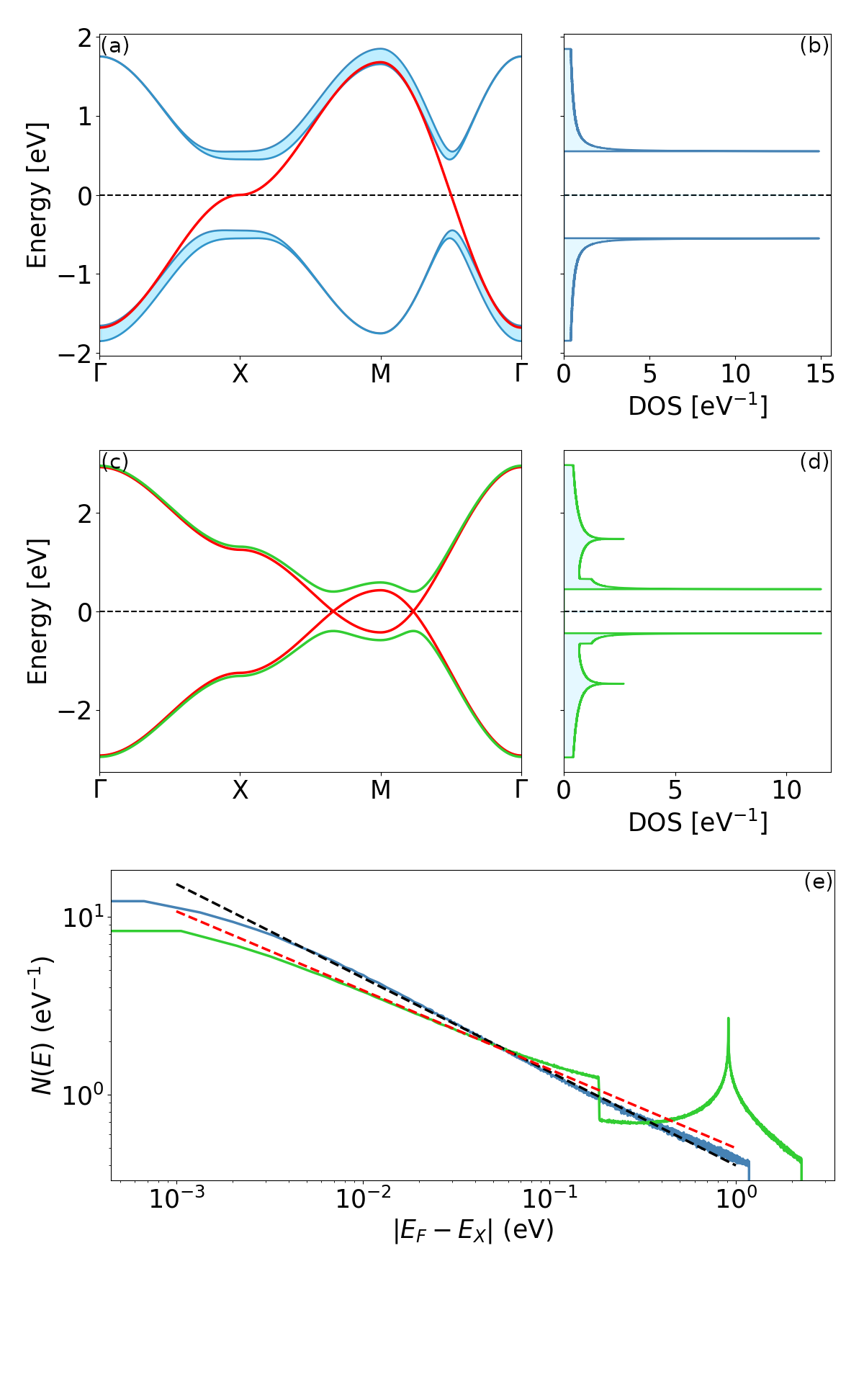}
\caption{{\bf Emergence of secondary VHSs} (a) $(\pi,\pi)$ Antiferromagnetic Hubbard model $(t'=t''=0)$ at half-filling with bare dispersion (red line) and gapped dispersion with mean-field gap parameter $\Delta= 0.5$~eV (blue curves). The width of the blue curve indicates spectral intensity. (b) The associated DOS. (c) Excitonic insulator model with bare dispersion (red line) and gapped dispersion with mean-field gap parameter $\Delta= 0.4$~eV (green curves) and (d) the resulting DOS. (e) Panels (b) and (d) are replotted on a {\it ln-ln} plot to show the high-order VHSs. The dashed black and red lines provide the reference slopes of -0.54 and -1/2, respectively.}
\label{fig:3b}
\end{figure}

{\it Secondary VHSs.$-$} The above analyses do not exhaust the possibilities for high-order VHSs. When a phase transition opens a gap in the electronic spectrum, the resulting subbands each develop secondary saddle-point VHSs, with corresponding new families of higher order VHSs having properties that can be quite different from the primary VHSs discussed above. Here we provide two examples of such secondary VHSs.  
Firstly, let us consider the DOS in the mean-field antiferromagnetic (AFM) phase in a pure Hubbard model ($t'=t''=0$), where the secondary VHS displays strong frustration.  Figures~\ref{fig:3b}(a) and (b) show that the DOS has a strong power-law divergence associated with quasi-1D nesting along the Brillouin zone diagonals, with enhanced flatness near $(\pi,0)$ and $(0,\pi)$.  Notably, when $t'$ is non-zero this evolves into a Mexican hat dispersion, with a local maximum at $(\pi/2,\pi/2)$, SM S-III.A.  With increased doping, the AFM gap closes near the point where the high-order VHS of the lower magnetic band crosses the fermi level, leading to a discontinuous loss of AFM order (see Ref.~\onlinecite{RSMstripe} and Fig.~17 of Ref.~\onlinecite{RM70}).  Such a scenario was recently seen experimentally\cite{Michon}, but the VHS interpretation was discarded because the feature was too intense to be a conventional logarithmic VHS \cite{Michon,Horio}.

Secondly, we investigate the case of an excitonic insulator where electron and hole pockets bind together to form avoided crossings at the Fermi level. While the electron and hole pockets can have different geometries or be shifted by a fixed $q$, for simplicity, we consider pockets of the same area and shape, with $q=0$. Adding a hybridization term to such a model leads to an avoided crossing with a `Mexican-hat' dispersion (Fig.~\ref{fig:3b}(c)). From Fig.~\ref{fig:3b}(d) it is seen that this dispersion leads to a highly characteristic DOS in either hybridized band. The logarithmic VHSs of the original bands at $\pm 1.5$eV remain unchanged by hybridization whereas the band-edge VHSs split into two VHSs- a step and a power-law VHS as one moves towards the band edge. The power-law saddle points are associated with an unusual Higgs-like one-dimensionality in the bands where they are flat along the brim of the Mexican hat but form an extremum along the radial $k$-direction away from $(\pi,\pi)$. Similar avoided crossings are found in TiSe$_2$\cite{singh2017} and often seen in topological insulators with a band inversion \cite{Bansil2016}.    

\begin{figure}[h!]
\centering
\includegraphics[width=0.99\columnwidth]{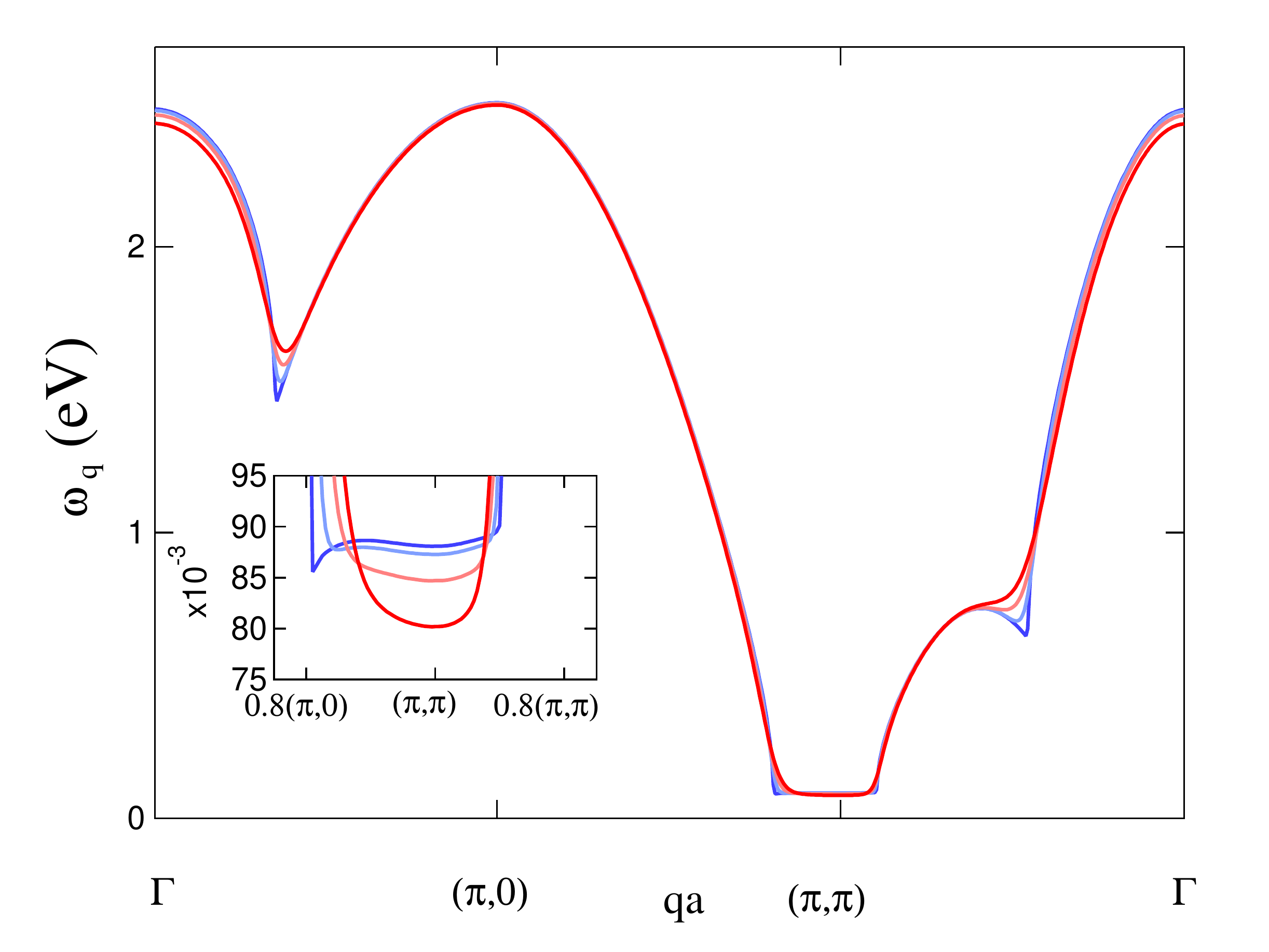}
\caption{{\bf Cuprate bosonic dispersion $\omega_q$.}    Dispersion at four different temperatures: T = 300 (blue line), 200 (pale blue), 100 (pink), and 10K (red). Inset: Blowup near band minimum.}
\label{fig:3c}
\end{figure}

We emphasize that these VHSs can be dubbed as Overhauser VHSs \cite{Over} since his model of charge density waves involves singular interactions on a 3D spherical Fermi surface. The circular VHSs would therefore provide a realistic 2D version of this effect. In this example, a flat band leads to strong frustration which can greatly lower the transition temperature. We find in Fig.~\ref{fig:3c}(e) that the secondary VHSs have distinct power laws ($p_V = 0.5, 0.54$) from the primary VHS ($p_V = 0.29$).  These exceptionally strong divergences satisfy the criteria required in high-order VHSs. The secondary high-order VHSs thus can be used as a signature of excitonic instabilities in materials. This is reasonable since the optical spectra (i.e. the joint DOS) of many semiconductors and insulators are dominated by prominent VHSs and an analysis of their associated dispersion geometries would ease the identification of excitonic states \cite{RSMdd, Liang4,JCP}.  Further details on both these secondary VHSs are found in SM Section S-III.

{\it High-order VHSs in Bosonic systems.$-$} Since VHSs occur in bosonic systems, one can ask the question if those VHSs can be of high-order. Here we demonstrate that bosonic VHSs carry features similar to the secondary VHSs discussed above. Reference~\onlinecite{MBMB} introduced the idea of a susceptibility density of states (SDOS) and showed its usefulness in mode-coupling theory and as a map of Fermi surface nesting. There has been recent interest in interpreting this susceptibility as a bosonic Green's function for electron-hole pairs\cite{Bosemet,spinon}.  We therefore consider $\chi_0(q,\omega)=\frac{1}{(\omega +\omega_{q0}-i\gamma_q)}-\frac{1}{(\omega-\omega_{q0}+i\gamma_q)}$, where $\omega_{q0}$ is a bosonic (electron-hole) frequency and $\gamma_q$ a damping rate. For $\omega\rightarrow 0$, $\chi_0$ becomes real and $\gamma_q\rightarrow 0$, so $\chi_0^{-1}=\omega_{q0}/2$ which gives the bosonic DOS up to a factor of 2. Coulomb interaction renormalizes $\omega_{q0}$ to $\omega_q=\omega_{q0}-2U$, where $U$ is the Hubbard interaction, Fig.~\ref{fig:3c} (see SM S-III.D). The dispersion in Fig.~\ref{fig:3c} looks unusual because the electronic susceptibility contains nonanalytic features at $T=0$ due to Fermi surface nesting, which also show up in Kohn anomalies of phonons \cite{RSMch}.  Interestingly, we find that both the Mexican hat and drumhead (flat-band) dispersions exist in Bosonic systems and give rise to high-order VHSs similar to the electronic case shown in Fig.~\ref{fig:3b} or (Fig.~2 of Ref.~\onlinecite{MBMB}). 

The bosonic high-order VHSs are particularly appealing in cuprates since they make the transition from commensurate AFM to incommensurate spin-density wave (SDW) highly anomalous. At the crossover point, any signs of the electronic order are lost, leading to an emergent spin-liquid phase. Since the AFM corresponds to what one expects from a Hubbard model (insensitive to shape of the Fermi surface), while the SDW is driven by Fermi surface nesting, it is appropriate to call this a Mott-Slater transition \cite{MBMB}, and the emergent spin-liquid phase suggests why it is so hard to explain cuprate superconductivity starting from the undoped insulator. A similar commensurate-incommensurate transition with hints of emergent spin-liquid behavior at the crossover has been observed for the 3D Hubbard model which has finite-temperature phase transitions \cite{KohnAn}.

Insight into the bosonic ring and drumhead dispersions can be gained from phononic dispersions. The electrons can be considered as moving in a quasi-static potential generated by the phonons, and a phonon soft-mode introduces a new component to the potential. The ring dispersion thus causes the electrons to move in a Mexican-hat dispersion which is a signature of the Jahn-Teller effect. For phonons, the Mexican hat dispersion is typically connected to a point of conical intersection, where several phonon modes are degenerate. This point typically lies above the brim of the Mexican hat, signaling that the high-symmetry point is unstable.  Notably, the resulting strong electron-phonon coupling leads to highly exotic physics, including breakdown of the Born-Oppenheimer approximation and, possibly, time crystals \cite{Bob1993}.

{\it Discussion.$-$}
While high-order VHSs are likely to play a prominent role in strongly-correlated materials and exotic superconductivity, a quick review of the various ways high-order VHSs arise in cuprates illustrates how complicated their role can be.  There is inter-VHS nesting responsible for the strong AFM effects and Mott physics; a bosonic high-order VHS that controls the transition from commensurate $(\pi,\pi)$ AFM to incommensurate SDW order, with an emergent spin-liquid phase \cite{MBMB}; and  crossovers, both at a critical $t'_c$ and at doping $x_{cross}$, from $(\pi,\pi)$ VHSs to $\Gamma$-centered VHSs that may be associated with the peak of the superconducting dome. 
\begin{figure}[h!]
\centering 
\includegraphics[width=0.5\textwidth]{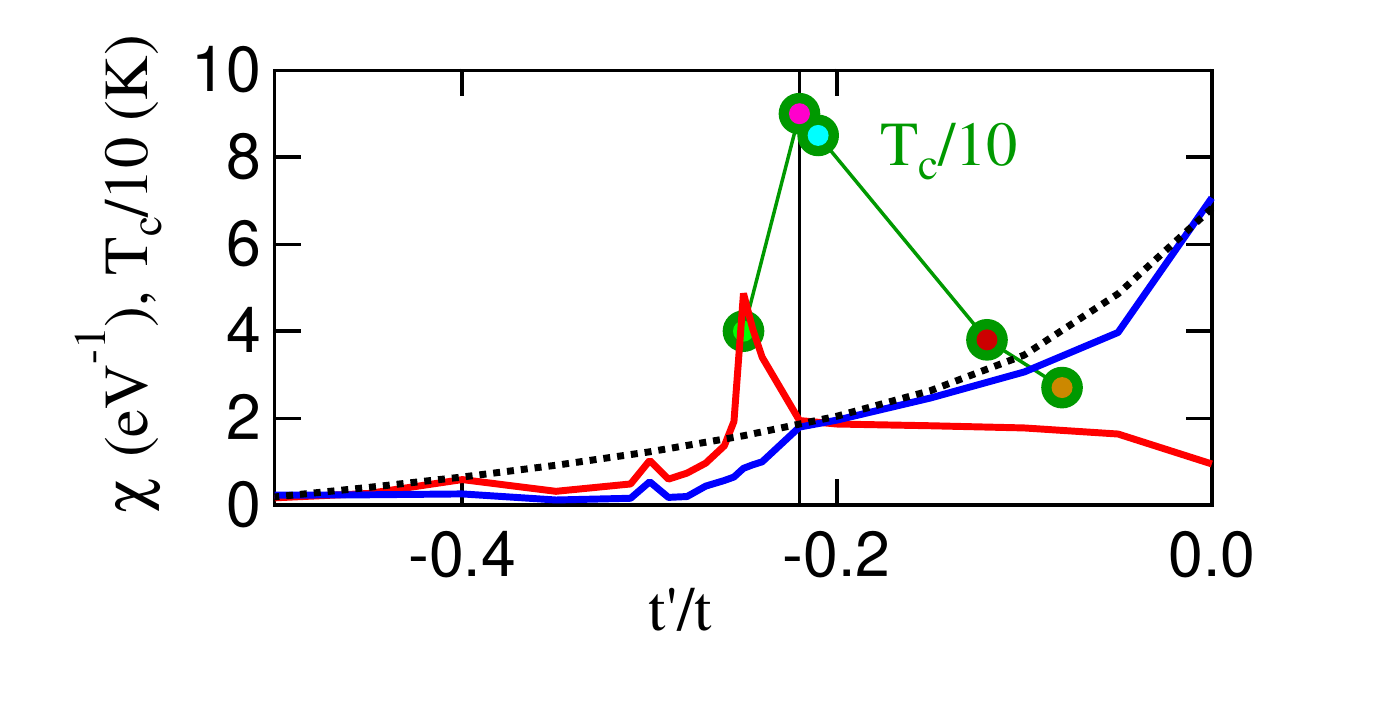}
\caption{ Competition between near-$(\pi,\pi)$ (blue line) and $\Gamma$-VHSs (red line), for $t''/t'$ = -0.5.   Black dotted line indicates analytic approximation for the former.  Black vertical line marks $t'_{cross}$.  Green line plots $T_c/10$ for monolayer cuprates of Fig.~2(b), using same color code for the dots.}
\label{fig:3} 
\end{figure}

To further illustrate the last point, in SM Section S-II.A we present susceptibility calculations of the VHS competition.  Figure~\ref{fig:3} shows the close connection between the crossover at $t'_c$ and optimal $T_c$ for the cuprates.  We see that for small $|t'|$ cuprate physics is dominated by AFM order associated with the $(\pi,\pi)$ VHS, but the hoVHS at $t'/t=-0.258$ drives a crossover at $t'_c/t=-0.22$ to the predominantly $\Gamma$ dominated VHS peak, while the highest superconducting $T_c$ of the Hg cuprates falls just at the crossover.  A similar crossover in each cuprate should occur as a function of doping at $x_{cross}<x_{VHS}$. 

This observation suggests possible explanations as to why $T_c$ in cuprates is so high.  For example, if high -$T_c$ superconductivity arises  by tipping the balance between the competing orders, favoring the $(\pi,\pi)$ instability, then superconductivity should be most important near $t'_C4,$ when the dominant instability crosses over from $(\pi,\pi)$ to $\Gamma$. In this case, $T_c$ should maximize near $x_{cross}$ and decrease with larger doping. Alternatively, superconductivity could actually benefit from the proximity of the strongly correlated and frustrated Mott phase with the weakly conducting Slater phase\cite{ABB}.  These scenarios provide a good description of the cuprates. For example, in Bi2212 $T_c$ is maximal near $x_{cross}$, and $\rightarrow 0$ near $x_{VHS}$\cite{Zhou}, where the pairing strength also vanishes \cite{TallStorey}.  
  
Additional issues include understanding the dual role of a VHS, in both increasing correlations via the peak in the DOS/susceptibility, and decreasing correlations by enhancing dielectric screening, and the role of secondary VHSs in driving/suppressing further instabilities. The bosonic high-order VHSs may have important relevance to Bose metals \cite{Bosemet} and spinon bands \cite{spinon}. There is an ongoing search for exotic phase transitions that do not fit into the conventional Moriya-Hertz-Millis model of quantum criticality, particularly when nontrivial emergent excitations arise near the quantum critical point \cite{Sachdev}. Thus, our finding that such an emergent phase can be driven by bosonic high-order VHSs may constitute the most direct evidence of the importance of high-order VHSs in correlated materials.

After years of debate on the significance of the VHSs in the cuprates, it is exciting to realize that they can be significantly more singular than previously imagined. The correlation between higher VHSs and higher superconducting $T_c$ we delineate here shows that high-order VHSs play a crucial role in high-$T_c$ cuprate superconductors.  However, the complicated nature of this correlation can be appreciated by noting that the superconducting $T_c$ does not usually maximize at the VHS doping $x_{VHS}$ but at a considerably lower doping. It will be interesting to see if similar complex intertwined orders are associated with high-order VHSs in other materials.

We now show that high-order VHSs exist if the electronic dimensions are smaller than lattice dimensions. During the early days of many-body perturbation theory, it was postulated that the effective electron dimensionality could be smaller than the crystal lattice dimension {\it i.e.} if a Fermi surface has flat parallel sections there would be good nesting, leading to quasi-1D behavior.  For example, the early high-T$_c$ superconductors such as the A15 compounds were assumed to be composed of three orthogonal interpenetrating chains of electrons \cite{Labbe}.  The three VHSs we have discussed in the cuprates illustrate this point.  The high-order VHSs of the bare dispersion remain point-like, but with power-law divergence, $p_V=0.29$, while the dispersion of the secondary AFM VHS is flat along an extended, curved line -- i.e., quasi-one dimensional, with $p_V=0.54$, and the bosonic VHS is flat over a finite area.  

The above results signal a substantial extension to the definition of a higher-order VHS\cite{Liang3}. Recall that a 2D saddle-point VHS corresponds to a point in $k$ space where the local dispersion has the form $ak_x^2-bk_y^2$, leading to a logarithmic peak in the DOS while a high-order VHS was defined as a point-like object where, for example, $b\rightarrow 0$. In contrast, here we consider the possibility that $b=0$ over an extended line segment.  This definition of high-order VHS is necessary, since the 2D materials must have a diverging VHS, and this divergence we identified is the only candidate. Such a definition can help to classify high-order VHSs in a wide range of flat-band materials, which are also characterized by having dispersions with flat line- or area-segments.

One final note on the connection between the higher-order VHSs and flat bands.  We have demonstrated that higher order VHSs are surrounded by a broad wake of 'enhanced-logarithmic' VHSs, that start out with a power-law divergence before crossing over to logarithmic, as seen in Fig.~\ref{fig:1vhs}(d).  On the other hand, in the study of flat bands, there is no consensus on how flat is flat: does a material count as a flat band if there is still a small dispersion?  We hypothesize that many flat band materials are associated with these enhanced-logarithmic VHSs, becoming flatter as they approach more closely to the higher-order VHSs.

\section*{Acknowledgements}     
We thank Adrian Feiguin for stimulating discussions. This work is supported by the US Department of Energy, Office of Science, Basic Energy Sciences grant number DE-FG02-07ER46352, and benefited from Northeastern University's Advanced Scientific Computation Center (ASCC) and the allocation of supercomputer time at NERSC through grant number DE-AC02-05CH11231.  The work at LANL was supported by the U.S. DOE NNSA under Cont. No. 89233218CNA000001 through the LANL LDRD Program and the CINT, a DOE BES user facility.  

\section*{Author contributions}
R.S.M., B.S., C.L., and A.B. all contributed to the research reported in this study and the writing of the manuscript.

\section*{Data Availability}
All data supporting the findings of this study are available from the corresponding authors upon request.

\section*{Additional information}
The authors declare no competing financial interests.

\end{document}

\title{
--- Supplementary Information ---\\*[1em]
High-$T_c$ superconductors as a New Playground for High-order Van Hove singularities and Flat-band Physics }

\author{Robert S. Markiewicz}
\email{r.markiewicz@northeastern.edu}
\affiliation{Department of Physics, Northeastern University, Boston, Massachusetts 02115, USA}

\author{Bahadur Singh}
\email{bahadur.singh@tifr.res.in}
\affiliation{Department of Condensed Matter Physics and Materials Science, Tata Institute of Fundamental Research, Colaba, Mumbai 400005, India}

\author{Christopher Lane}
\affiliation{Theoretical Division, Los Alamos National Laboratory, Los Alamos, New Mexico 87545, USA}
\affiliation{Center for Integrated Nanotechnologies, Los Alamos National Laboratory, Los Alamos, New Mexico 87545, USA}

\author{Arun Bansil}
\affiliation{Department of Physics, Northeastern University, Boston, Massachusetts 02115, USA}

\maketitle

\beginsupplement
\section{Optimizing $t'$ values}
\begin{figure}
\leavevmode
\rotatebox{0}{\scalebox{0.50}{\includegraphics{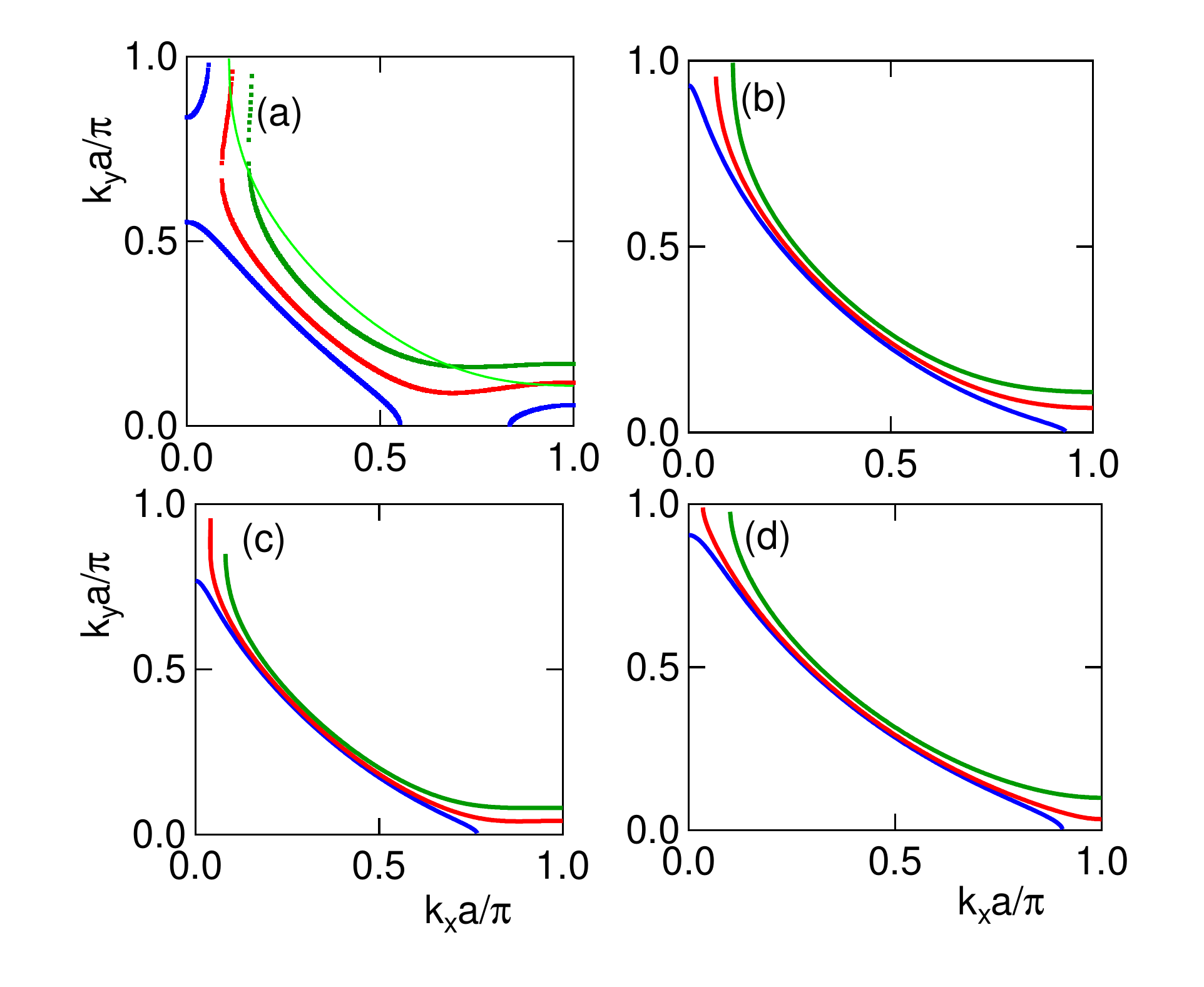}}}
\vskip0.5cm
\caption{
{\bf Proposed Fermi surface models of Tl- (a,b) and Bi-cuprates (c,d)}, for $t'/t$ = -.32 (a), -0.22 (b), -0.26 (c), and -0.17 (d).}
\label{fig:6}
\end{figure}

The fermi surfaces of the antiferromagnetic, electron-doped cuprates\cite{nparm} have become an iconic image in physics -- hole pockets around $(0,\pi)$ and $(\pi,0)$, electron pocket near $(\pi/2,\pi/2)$, separated by hot-spots. Yet modeling that structure initially proved difficult.  Indeed, in the commonly used $t-t'$ model the hole pocket never appears.  However, Andersen\cite{OKA} provided chemical arguments that cuprates are better described by a $t-t'-t''$ model with the special value $t''=-t'/2$.  By adding precisely this value of $t''$, the hole pocket was readily produced\cite{Kusko}.  Similarly, for hole doped cuprates, modeling them with a $t-t'$ model with different $t'$ for different families of cuprates is unsatisfactory, leading to too strongly nested fermi surfaces for larger $t'$, whereas adding $t''=-t'/2$ produces reasonable results\cite{MBMB}.

While we adopt Andersen's ratio $t''/t'=-1/2$ in this work, our philosophy of using the $t-t'-t''$ model is different from that of Pavarini {\it et al}\cite{PavOK} (PA). PA downfold the DFT dispersions of several bands to a single band representing the cuprates, then fit the band to a single parameter $r$, which is approximately equal to $t'/t$.  We find that improved fits to the full dispersion require more than three hopping parameters,\cite{RSM1b} which can be used to calculate the susceptibility, and thence the self-energies and phase diagrams, either at random-phase approximation\cite{Das} or with mode-coupling corrections\cite{MBMB}. Then, we introduce a {\it reference family} -- the $t-t'-t''$ model for a given cuprate whose susceptibility most closely matches that of the many-$t$ model as a functon of doping and temperature.  In this way, we find that the family with $t''=-0.5t'$ provides a reasonable model for cuprates, where each cuprate is associated with a particular value of $t'$, and one can easily explore parameter space where no cuprates are found.\cite{MBMB}

However, by comparison of individual materials to experiment, we find that the PA values $t'/t=\alpha r$, $\alpha = 1$, tend to overestimate $x_{VHS}$.  Here, we focus on a comparison of Tl2201 and Bi2201, Fig.~\ref{fig:6}, where experimental data on the fermi surfaces are available\cite{Damasc,deHuss,Valla}. In Fig.~\ref{fig:6}, we compare fermi surfaces assuming the PA values of $t'$ ($\alpha =1$) for Tl2201 (frame (a), $t'/t$ = -0.32) and Bi2201 (frame (c), $t'/t$ = -0.26) with smaller values of $\alpha =2/3$ for Tl2201 (frame (b), $t'/t$ = -0.22) and Bi2201 (frame (d), $t'/t$ = -0.17).  Clearly, frame (a) bears little resemblance to the known fermi surface of Ti2201\cite{Damasc,deHuss}, whereas frame (b) provides a much better match.  In contrast, for Bi2201, the $\alpha=1$ result comes much closer to the fermi surface seen in ARPES\cite{Valla}, while the theoretical $x_{VHS}=0.50$ is close to the experimental $0.45$ (contrast $x_{VHS}=0.29$ for frame (d)).  Moreover, the area of the experimental Tl2201 fermi surfaces, not at the VHS, corresponds to a doping $x$ = 0.26\cite{Damasc} or 0.24\cite{deHuss}, to be compared to the dark green curves in frames (a) and (b), both calculated at $x$ = 0.25.  It can be seen that the $t'=-0.22$ fermi surface (for convenience reproduced as a light green line in frame (a)) approaches much closer to $(\pi,0)$, in better agreement with experiment\cite{Damasc,deHuss}.  The corresponding theoretical values of $x_{VHS}$ are 0.563 (frame (a)) and 0.398 (frame (b)).  This suggests that the PA $t'$ values are misordered, and that Bi-cuprates have the largest value of $t'$.  The simplest correction is that $\alpha$ = 1 for Bi2201 and = 2/3 for the other cuprates.  This corrects the Tl2201 value and brings LSCO into good agreement with Ref.~\onlinecite{MBMB}, while keeping the Tl and Hg values close to each other.

The finding that Bi2201 has the largest $t'$ value could explain several puzzles of the Bi cuprates.  First, the small $T_c$ in Bi2201 could be due to too large a value of $t'$, causing the VHS to fall either at the hoVHS or perhaps just off the $(\pi,\pi)$ plateau.  Moreover, the large enhancement of $T_c$ in Bi2212 follows since bilayer splitting reduces the $t'$ of the antibonding band into a more appropriate range of $t'$, with $x_{VHS}\sim 0.32$.\cite{RSM1b} [See discussion below, near Eq.~\ref{eq:4b}.]

We note that with the $\alpha\sim$~2/3 correction, superconductivity in cuprates is optimized for a band structure close to that of the hoVHS.  This also means that high $T_c$ superconductivity is confined on the $(\pi,\pi)$-plateau.  [For comparison purposes, we refer the reader to our original version, ArXiv:2105.04546.v1, Fig. 1, to see the comparison of VHS and $T_c$ for the original PA parameters.]

\section{Van Hove dichotomy}
\subsection{At the VHS, vs $t'$}

In the main text, we have considered VHSs with diverging $q=0$ susceptibility. However, the VHS also has a diverging susceptibility at a second wave momentum $Q$ near $(\pi,\pi)$ as well. This is resolved in Fig.~4 of the main text, where the variation of $t'$ shifts the dominant susceptibility peak from $(\pi,\pi)$ to $\Gamma=(0,0)$ as $-t'$ increases. Similar effects arise as doping $x$ increases from 0.  This dichotomy is very relevant for cuprate physics since only the VHS near $(\pi,\pi)$ is correlated with the pseudogap crossover temperature, $T_Q\simeq T_{pg}$. We now demonstrate how these two VHSs compete and evolve with $t'$ and $x$.

\begin{figure}[h!]
\centering 
\includegraphics[width=0.5\textwidth]{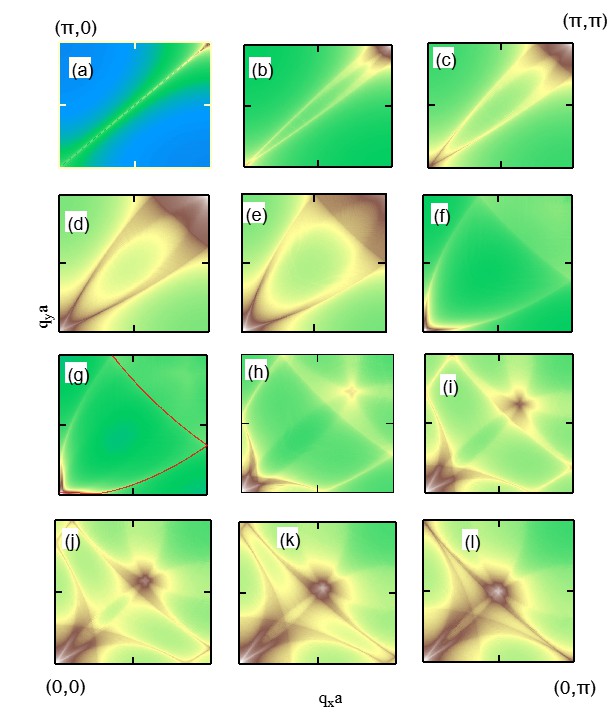}
\caption{Susceptibility maps at saddle-point VHS for $t'/t$ = (a) 0.0, (b) -0.05, (c) -0.10, (d) -0.15, (e) -0.20, (f) -0.25, (g) -0.258 [Bi2201], (h) -0.30, (i) -0.35, (j) -0.40, (k) -0.45, and (l) -0.50.   Blue and white identify minimum and maximum intensities, respectively.}
\label{fig:2} 
\end{figure}

To capture the dichotomy, we calculate the full DFT-Lindhard susceptibility, 
\begin{equation}
\chi_0(q,\omega)=-\sum_k\frac{f(\epsilon_k)-f(\epsilon_{k+q})}{\omega -\epsilon_{k+q}+\epsilon_k}
\label{eq:1}
\end{equation}  
The real part of $\chi_0(q,\omega =0)$ contains two contributions: the diverging peaks associated with VHS nesting\cite{RiSc} plus a folded ($q=2k_F$) map of the Fermi surface, whose intensity quantifies the strength of Fermi surface nesting.  The susceptibility is not confined to the Fermi surface, but contains a substantial bulk contribution which can favor certain $k_F$ values, or even shift the peak to a near-by commensurate value, necessitating a careful numerical evaluation of $\chi_0$.\cite{MLSB}
   
Figure~\ref{fig:2} shows the evolution of the susceptibility map at the VHS doping as $t'$ is varied, keeping $t''=-0.5t'$. In \Fref{fig:2}(g), we show the Fermi surface nesting curve $q=2k_F$ (red line) over part of the first Brillouin zone. This clearly identifies that the nonanalytic features of the susceptibility are associated with Fermi-surface nesting.  It can be seen that for each $|t'|$, there are susceptibility divergences at $\Gamma$ and  $(\pi,\pi)$, due to intra- and inter-VHS scattering, respectively.   However for small $|t'|$, the more intense VHS feature is at $(\pi,\pi)$ whereas for larger $|t'|$, the peak at $\Gamma$ becomes most intense.  From Fig.~4, we note that the crossover occurs near $t'=-0.22t$, pinned by the high-order VHS peak.  For intermediate values of $|t'|$, there is an additional peak at $(\pi-\delta,\pi-\delta)$ (see \Fref{fig:2}(h) for example). 
 
To understand the $(\pi,\pi)$ VHS evolution, we recall that for the pure Hubbard model ($t'=0$), the $(\pi,\pi)$-susceptibility has a $log^2T$ divergence that for finite $t'$ evolves to 
$$\chi((\pi,\pi),0)\sim log(T)log(T_X),$$ where $T_X=max\{T,T_{eh}\}$ and $k_BT_{eh}=|t'|$.\cite{VHS2} The dotted black line in Fig.~4 is proportional to $ln(|t'|)$ (with an effective $t'$ at $t'=0$, to account for the finite $k$-mesh), showing that this is a good approximation over a broad range of $t'$, even in the presence of $t''\ne 0$.

Near high symmetry points the susceptibility peak can deviate from the nesting map.  This can be understood as follows, taking the symmetry point $(\pi,\pi)$ as an example.  While the susceptibility at $\Gamma$ is constrained to be a Fermi surface effect, for general $q$ there can be a strong contribution to $\chi_0$ from states far from $E_F$, leading to peaks with very broad tails.  These tails play an important role near high symmetry points, when multiple Fermi surface sections approach the symmetry point.  Figure~\ref{fig:2} illustrates this effect for the $(\pi,\pi)$ plateau.   For larger $|t'|$, Figs. \ref{fig:2}(f)-(l), the weight is concentrated near the nesting map, $q=2k_F$.  However, as $|t'|$ decreases, overlap of the bulk spectral weight rapidly shifts the peak to exactly $(\pi,\pi)$, as the plateau shrinks down to a point for $t'=0$ (Fig. \ref{fig:2}(a)).  Similar overlap effects explain why the susceptibility peaks near $\Gamma$ can shift away from the nesting map.

The pileup of states at $(\pi,\pi)$ for small $|t'|$ has two consequences- first, a prominent commensurate-incommensurate transition\cite{MLSB}, and second, the rapid decrease of intensity at $(\pi,\pi)$ with increasing $|t'|$ seen in Fig.~4.  We note that the Fermi surfaces responsible for the $q\sim (\pi,\pi)$ plateau come from $k_F\sim (\pi/2,\pi/2)$ -- i.e., far from the VHS and hence not sensitive to $t''$.  Indeed, the $t'$-dependence of the $(\pi,\pi)$ plateau area, \Fref{fig:2}, is qualitatively similar to the doping dependence of the plateau area at fixed $t'$, \Fref{fig:1}.

\subsection{Tuning away from the VHS at fixed $t'$}
\begin{figure}[h!]
\centering 
\includegraphics[width=0.5\textwidth]{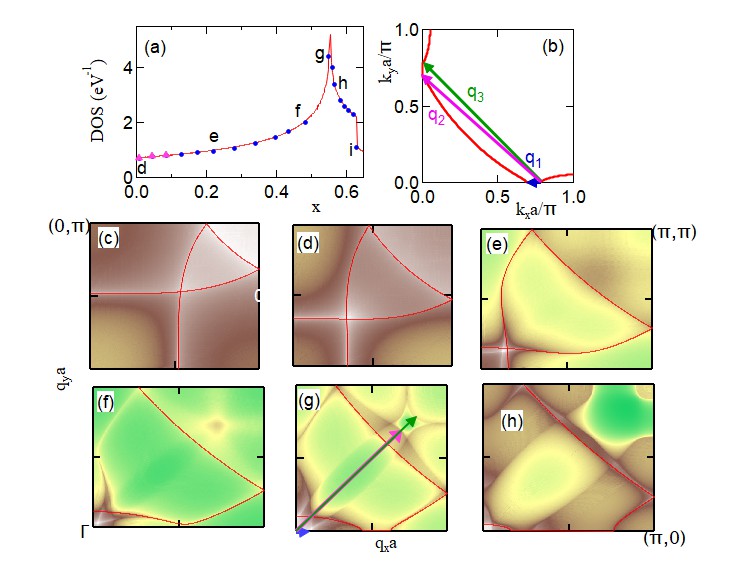}
\caption{{\bf Susceptibility maps showing evolution of nesting lines with doping for $t'/t=-0.3$, $t''/t'=-0.5.$} 
(a) DOS, showing doping of each susceptibility map. (b) Fermi surface for frame (g) showing three inter-surface nesting vectors.  (c-h) Susceptibility maps, with $q=2k_F$ nesting surface superposed as a red line; in frames (f-h), one branch is omitted to emphasize details of the nesting maps. Color scheme same as \Fref{fig:2}.  In frames (c-h), the $(\pi,\pi)$-plateau is defined as the area enclosed by the folded Fermi surface nearest to $(\pi,\pi).$  Arrows in (g) match those in (b).}
\label{fig:1}
\end{figure}

While the $t'$ dependence of the VHS susceptibility is easy to calculate, since only the doping $x_{VHS}$ is involved, the doping dependence at fixed $t'$ (i.e., for a particular cuprate) is often more important.  This is more complex since, for $x \ne x_{VHS}$, each VHS evolves into a finite susceptibility peak with different doping dependence.  We illustrate this evolution  in Figure~\ref{fig:1} for a typical $t'/t = -0.3$.  The susceptibility has a number of nonanalytic peaks, associated with Fermi surface nesting.  To illustrate this, we include superposed maps of the main barrel Fermi surfaces, doubled and folded back to the first Brillouin zone (BZ), to highlight the $q=2k_F$ nesting features.\cite{MLSB}  Additional features in frames (f-h) are associated with non-$2k_F$ nesting,\cite{MLSB,MGu}, e.g., inter-Fermi surface pocket nesting as illustrated in frame (b) for the doping of frame (g).  

To better understand the structure in these susceptibility maps, we briefly review the evolution of the Fermi surface, and its two topological transitions\cite{MLSB,MGu}.  Figure~\ref{fig:1}(a) shows the corresponding DOS, with points at which the susceptibility maps were sampled.  The DOS has two VHSs -- a logarithmic peak where a pocket pinches off near $(\pi,0)$, frame (f), and a step down when the pocket disappears with further hole doping, frame (h).  Due to the folding, the pockets appear near $\Gamma$ in the susceptibility maps.

For present purposes, our main result is to see how the VHS competition plays out as the Fermi level is shifted away from $E_{VHS}$.  As doping $x$ increases, the dominant nesting vector shifts away from $(\pi,\pi-\delta)$ at low doping (violet triangles in frame (a)) to $(\delta',\delta')$, or antinodal nesting (ANN), at higher doping (blue dots), near $x_{cross}\sim 0.12$.   Experimentally, the peak at $(\pi,\pi-\delta)$ is associated with an incommensurate SDW, while the ANN peak drives a CDW instability.  As the VHS peak is approached, the ANN peak flows to $\Gamma$, $\delta'\rightarrow 0$ at the VHS, while the peak at $(\pi,\pi-\delta)$ evolves to a weak peak at $(\pi-\delta'',\pi-\delta'')$ at the VHS, green circle in \Fref{fig:1}(f). Thus we have demonstrated that this competition is controlled by the competition between intra-VHS scattering near $\Gamma$ and the inter-VHS scattering near $(\pi,\pi)$.
  That is, the VHSs are important not only at $x_{VHS}$, but influence the dominant susceptibility over a wide doping range.  

The peak crossover seen at $x_{cross}=0.12$ in \Fref{fig:1} is reflected in a similar crossover of the VHS peaks, Fig.~4, at $t'_{cross}\sim -0.22$, for $t''=-0.5t'$.  For all $t'>t'_{cross}$ there is no $x_{cross}$ and the $(\pi,\pi)$-VHS is always dominant, while for $t'<t'_{cross}$ there is always an $x_{cross}$ where the influence of the $(\pi,\pi)$-VHS is lost.  Note that in the example of \Fref{fig:1}, $x_{cross}=0.12 << x_{VHS}=0.54$.

\section{More on secondary and bosonic VHSs}
\subsection{Many-body perturbation theory background}
The calculations of this paper are based on a many-body perturbation theory (MBPT) of the cuprates\cite{Das,MBMB}, using tight-binding approximations to DFT dispersions.  At the time these were developed, it was unclear whether DFT could capture the gapped physics of strongly correlated materials like cuprates, and it remains a useful tool for understanding the newer DFT results.  The theory implements Hedin's scheme\cite{Hedin}, wherein MBPT can be formally solved exactly by taking the lowest order perturbation theory results for the electron Green's function $G_0$, self energy $\Sigma_0$, susceptibility $\chi_0$, and vertex correction $\Gamma_0$, and replacing them by the fully dressed functions, $G_0\rightarrow G$, etc., wherein the dressed functions must be simultaneously found self-consistently.  

In Ref.~\onlinecite{Das}, an approximate $G_ZW_Z$ scheme was developed for solving these equations in the limit where vertex corrections are ignored, $\Gamma\rightarrow 1$ -- a form of GW approximation\cite{Hedin}.  The $G_ZW_Z$ scheme successfully captured the splitting of the electronic dispersion into a low energy branch of coherently dressed quasiparticles and a high energy incoherent background, separated by an almost discontinuous step in the dispersion at an intermediate energy -- the waterfall effect.  For present purposes, the key result is that the incoherent part of the spectrum can be ignored for the low energy physics, and the coherent part can be approximated as the DFT dispersion renormalized by a constant factor $Z<1$.  The resulting gap equation can then be approximated by either the random-phase approximation (RPA) or the Hartree-Fock gap equation for the renormalized quasiparticles in terms of a renormalized Hubbard interaction $U$.  Since DFT calculations do not capture the renormalization, we further simplify our calculation by setting $Z\rightarrow 1$, which does not change the fermi surface or the shape of the DOS, except for an overall factor of $Z$.  These results are discussed in Section S-III.B.

Reference~\onlinecite{MBMB} extended the quasiparticle part of Ref.~\onlinecite{Das} by including a vertex correction to account for mode coupling.  This is important to account for the dominance of short-range order in cuprates, produced by a combination of Mermin-Wagner\cite{MW} physics and frustration from competition between different electronic instabilities -- an electronic analog of McMillan's bosonic entropy in srongly-correlated charge density waves\cite{McMill}.  In defining the mode coupling, a natural parameter arose -- the inverse-susceptibility DOS\cite{MBMB}.  Here we extend the analogy between electronic spin-density waves and phonon-induced charge density waves by rewriting the electronic susceptibility in terms of the Green's function of an electronic boson, which represents an electron-hole pair or parexciton.  This idea is developed in Section. S-III.D.

\subsection{AFM VHSs}
To our knowledge, the self-consistent Hartree-Foch calculation was the first to demonstrate the important role of the VHS of the lower AFM band in causing the first-order collapse of the AFM order.\cite{RSMstripe}  For arbitrary hopping parameters, AFM order splits the band into lower and upper subbands, with dispersion
\begin{equation}
\epsilon_{\pm}=\epsilon_{0+}\pm\sqrt{\epsilon_{0-}^2+\Delta^2},
\label{eq:14}
\end{equation}
where $\epsilon_{0-}=-2t(c_x+c_y)$, $\epsilon_{0+}=-4t'c_xc_y-2t''(c_{2x}+c_{2y})$, and the gap $\Delta$ must be found self-consistently as a function of doping.  Physically, the DOS of the lower band $\epsilon_-$ is more important, although for the Hubbard model ($t'=t''=0$) discussed in the text both bands have the same dispersion.  Since the AFM order is sensitive to the exact value of $U$, we use the quaiparticle-GW parameters\cite{Das} $U=3t$ and $Z=0.5$, following Ref.~\onlinecite{gapsend}.

\begin{figure}[h!]
\centering 
\includegraphics[width=0.5\textwidth]{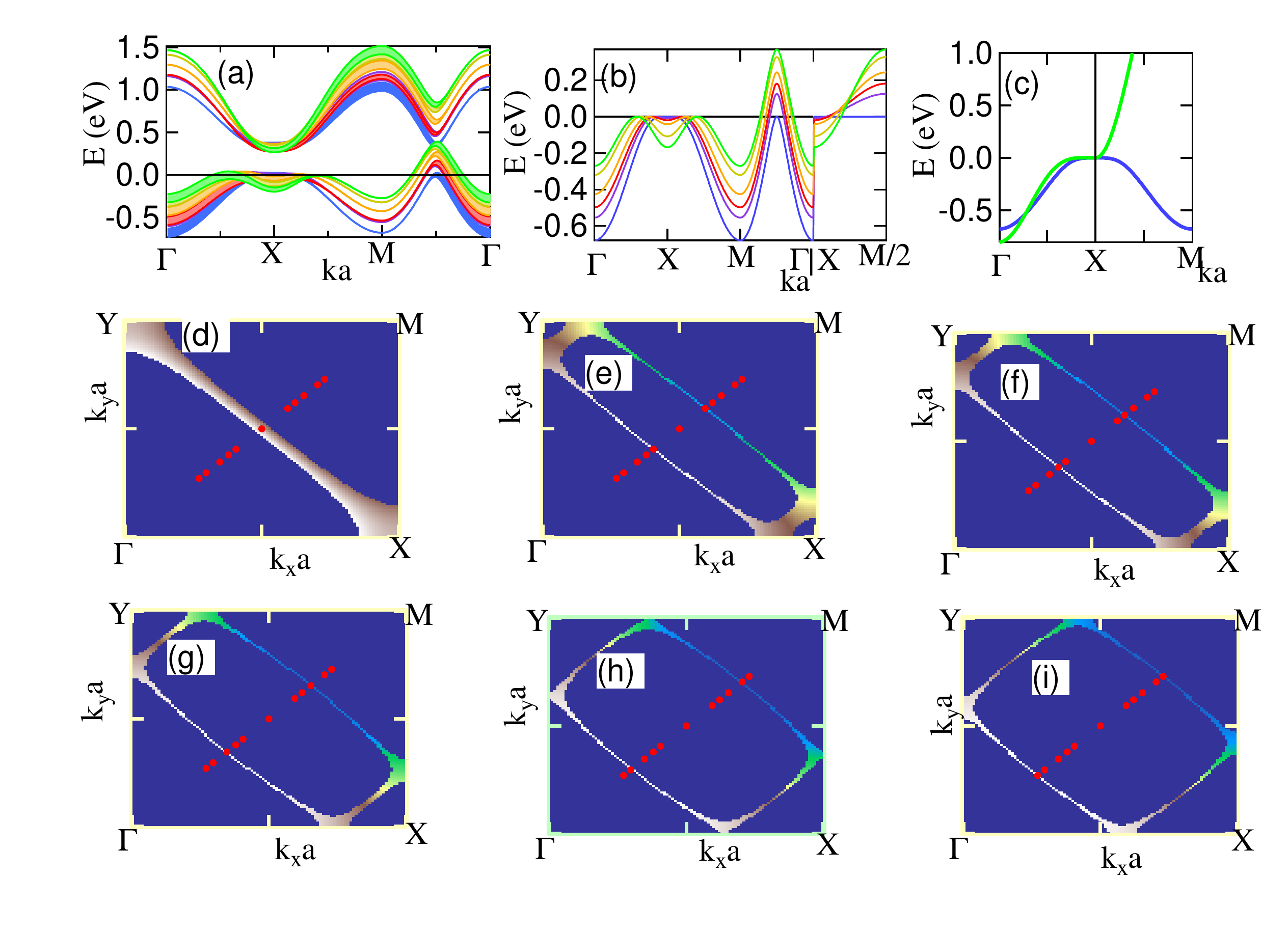}
\caption{{\bf (a-c) AFM dispersion maps showing evolution of VHS energy (at $E=0$) with $t'$} for $t''/t'=-0.5,$ with (a) and without (b,c) spectral weight correction.  Colors in (a,b) and VHS energy surface in (d-i) correspond to $t'/t = $ (d) 0 (blue, $\Delta = 0.283$ eV), (e) -0.08 (violet, 0.2828), (f) -0.12 (red, 0.2829), (g) -0.17 (orange, 0.3044), (h) -0.26 (gold, 0.3482), and (i) -0.32 (light green, 0.3674).  Frame (b): corresponding dispersions along AFM zone boundary, $X\rightarrow M/2$,  showing characteristic $C+Ac_x^2$ form, with $A\propto t'$.  (c): contrast of hoVHS dispersions for NM (light green, from Fig.~1(b)) and AFM (blue, from (a)).}
\label{fig:A9}
\end{figure}

We here discuss Eq.~\ref{eq:14} for more general dispersions, recalling that in general $\epsilon(k)_{0\pm}=(\epsilon(k)\pm\epsilon(k+Q)/2$, where $Q$ is the AFM wave number, in our case $Q=(\pi,\pi)/a$, and we explicitly show the lattice constant $a$.  We find that the AFM VHS evolution is controlled by the dispersion along the AFM zone boundary (ZB), which corresponds to $k_x+k_y=\pi/a$, plus symmetry-related lines.  In Fig.~\ref{fig:A9}, we show how the VHS evolves as a function of $t'$ for the self-consistent gap parameter, $\Delta=US$,\cite{gapsend}, where the value of the dimensionless magnetization $S=<n_{\uparrow}- n_{\downarrow}>/2$ are listed in the figure caption.  Along the AFM ZB $\epsilon(k)_{0}$ is identically zero, for arbitrary dispersion, and Eq.~\ref{eq:14} reduces to $\epsilon_{\pm}=\epsilon_{0+}\pm\Delta$, while along the AFM ZB $\epsilon_{0+}$ simplifies, becoming $\epsilon_{0+}=4t'c_x^2-4t''c_{2x}=4(t'-2t'')c_x^2+4t''$ for the $t-t'-t''$ model.  Thus for all $t-t'-t''$ models, the lower AFM band has an identical (4-sided) Mexican-hat minimum, 
\begin{equation}
\epsilon_-=C+Ac_x^2,
\label{eq:15}
\end{equation} 
where $C=4t''-\Delta$ and $A=4(t'-2t'')$ are constants. This behavior is clearly illustrated in Fig.~\ref{fig:A9}(b), where we plot half of the AFM ZB dispersion, $X\rightarrow M/2 (\rightarrow Y)$.  

This simple evolution along the AFM ZB leads to a universal behavior for the VHSs, Fig.~\ref{fig:A9}.  In an ordered phase, the dispersion in the superlattice Brillouin zone must be unfolded into the primitive Brillouin zone for comparison to experimental angle-resolved photoemission (ARPES) dispersions.  This process leads to a modulation of the spectral weight by a coherence factor, as discussed in conjunction with Fig.~2(b). Figure~\ref{fig:A9}(a) shows the corresponding dispersions for the AFM bands for 6 different values of $t'/t$, covering the range of the cuprates, with the width of the dispersion line being proportional to the spectral intensity.  This plot is repeated for the lower magnetic band in Fig.~\ref{fig:A9}(b) without the spectral weight for ease in identifying the VHS features.  For each curve, the fermi energy is adjusted so the VHS is at the fermi level, $E=0$.  From these figures, it is clear that the VHS is associated with the top of the dispersion along the lines $\Gamma\rightarrow X$ and $X\rightarrow M$.  [Parenthetically, we note that this is a good method for finding the energy of the VHS peak. Fig.~\ref{fig:A9}(h) shows how a small offset in $E_f$ distorts the corresponding fermi surface.]  

Having found the fermi energy of the VHS peak, Figs.~\ref{fig:A9}(d-i) show the six corresponding fermi surfaces.  In each frame, we plot the spectral weight for all data points within 5~meV of the fermi level, with dark blue corresponding to zero weight, and white to weight = 1.  As a check, the red dots correspond to the points where the dispersions cross the fermi level along the line $M\rightarrow \Gamma$ in frame (b).  The resulting pattern is remarkably simple: for $t'=0$, there is no dispersion along the AFM ZB, so the fermi surface lies along this line, spreading out near $X$ and $Y$.  When $t'\ne 0$, the ZB develops a dispersion, and the VHS fermi surface evolves into a box with the long sides split away symmetrically from the ZB, with the splitting increasing with $|t'|$.  While for $t'=0$, the DOS has a power-law divergence at the VHS, for $t'\ne 0$ one finds an enhanced logarithmic divergence\cite{gapsend}.

Since $\epsilon_{0-}$ universally vanishes on the AFM zone boundary, can we say something similar about $\epsilon_{0+}$ beyond the $t-t'-t''$ model?  Remarkably, we can.  In Section IV.B, we show that the tight-binding expansion including arbitrary higher order hopping in a 1-band model of cuprates can be rewritten as a Taylor series with arbitrary powers of $X=(c_x+c_y)$ and $Y=(c_x-c_y)$.  Along the AFM zone boundary, $X=0$ and $Y=2c_x$.  Thus, including more distant hopping can change Eq.~\ref{eq:15} to an arbitrary polynomial in $c_x^2$, where only even powers are allowed to preserve the $x-y$ symmetry of the lattice. It is likely that these extra terms will lead to more complicated shapes of VHS fermi surfaces.

\subsection{Excitonic Insulators}
While an exciton is usually considered to be a photoinduced excited state, an interesting situation arises if the exciton binding energy $E_x$ exceeds the gap between the conduction and valence bands from which it is formed.\cite{Kohn}  Then, the situation resembles a Cooper pair in a superconductor -- the assumed ground state is unstable and the true ground state is a condensate of excitons, or an excitonic insulator.  While the resulting formalism is similar to that of a superconductor, there is an important difference.  In a superconductor, the hole band is just the electron band reflected about the fermi energy, so the bands are always perfectly nested.  In the excitonic insulator the conduction and valence bands are independent, could fall at different $k$-points, and need not be well nested, all of which can affect the insulator's stability.  

In the text, we explored only the simplest case of perfect nesting at the same $k$ point.  To relate this problem to the present results, we assumed that the valence band $E_v$ is represented by the same $t-t'-t''$ model, and to ensure perfect nesting, we assumed that the shifted (by $E_x$) conduction band $E_c$ is just the mirror image of the valence band.  This leads to excitonic insulator bands of the same form as Eq.~\ref{eq:14}, but with $\epsilon_{0\pm}=(E_v\pm E_c)/2$, or for our choice of bands $\epsilon_{0+}=0$, $\epsilon_{0-}=E_v-\epsilon_f$, where $\epsilon_f$ is the fermi energy.

The same model would also describe the bands in a superconductor, except that in the cuprates the gap has d-wave symmetry rather than being constant.  We also note that similar phenomena may be found in topological nodal line semimetals\cite{NLSM1,NLSM2}, which have a closed nodal line of band degeneracy in the bulk and a sheet of drumhead surface states on the sample surface.  However, in general both the loop and the surface states can be strongly dispersing and it is not clear if a recipe for producing flat bands can be found.

\subsection{bosonic DOS}
In this Subsection, we discuss the inverse susceptibility DOS (ISDOS) introduced in Ref.~\onlinecite{MBMB}.  However, to put it in more familiar terms, we formally redefine the susceptibility in terms of a bosonic Green's function, in which case the ISDOS is simply the resulting bosonic DOS, and the resulting hoVHS peaks quantify McMillan's ideas of bosonic entropy driving strongly coupled CDW physics\cite{McMill}.  For the RPA, a magnetic instability arises from Dyson's equation, $\chi^{-1} =\chi_0^{-1}-U$, when the dressed susceptibility $\chi$ diverges.  In terms of the bosonic model, this should be a soft-mode instability, $\omega_q=\omega_{q0}-2U\rightarrow 0$.  However, it was found\cite{MBMB} that the RPA breaks down in the presence of strong mode coupling.  Specifically, the bosonic DOS quantifies the number of competing modes, adding a correction to the denominator of $\chi$ which slows its divergence.  

By reviewing the susceptibility maps in Fig.~\ref{fig:1}, we can understand what is happening in Fig.~3.  We focus on Fig.~\ref{fig:1}(c), where the effect is most clearly seen.  The susceptibility contains a nesting map -- a folded image of the fermi surface (red line) where the intensity represents the strength of fermi surface nesting for each point of the fermi surface.  In addition there is a $(\pi,\pi)$ plateau, the roughly diamond-shaped region of the susceptibility map centered on $(\pi,\pi)$ and extending to the fermi surface map on all sides (bright white area in Fig.~\ref{fig:1}(c).   Due to the fermi function cutoff, the susceptibility is much stronger on the plateau than off of it.  Moreover, as the plateau area decreases (lower doping or smaller $|t'|$), the tails of the VHS peaks overlap and push the suscepibility peak to exactly $(\pi,\pi)$ -- away from conventional fermi surface nesting.     

Thus, for a given material ($t'$), at low doping the susceptibility peaks at the commensurate value $(\pi,\pi)$ independent of doping.  As doping increases, the $(\pi,\pi)$ plateau expands, the peak height decreases, and at some point $(\pi,\pi)$ turns into a local minimum with the maximum susceptibility moved to an incommensurate $q$ on the ring of fermi surface nesting.  At the crossover the susceptibility becomes very flat on the plateau -- the coefficient of the $k^2$ term in a $k\cdot p$ expansion must vanish -- and the area of the flat band in Fig.~3 is approximately the area of the $(\pi,\pi)$-plateau.  Moreover, we have adjusted $t'=-0.267t$ (with $t''=0$) in Fig.~3 so that $\omega_q$ (as defined at RPA level) softens almost to zero as $T\rightarrow 0.$  However, due to the large bosonic DOS, the mode coupling correction overwhelms the mode softening, so that the correlation length remains very small even as $T\rightarrow 0$ leaving a spin liquid phase surrounded by commensurate or incommensurate phases with much larger correlation lengths.\cite{MBMB}  Reference~\onlinecite{MBMB} noted an interesting connection between this commensurate-incommensurate transition and the excited-state quantum phase transitions in the spectrum-generating algebras of nuclear theory.\cite{Iach}

\section{Improved tight-binding model}
\subsection{New $ln^2T$ VHSs}
In this Section, we propose a further generalization of higher-order VHSs, in which the faster-than-logarithmic divergence is not in the DOS, but in a competing, finite-$q$ susceptibility.  At present, the only known example is the original Hubbard model, $t'=t''=0$, with a $ln^2(T)$ $q=(\pi,\pi)$ susceptibility divergence.  Here, we demonstrate that the Hubbard model is not unique by generating a family of dispersions with this property.  As a byproduct, we demonstrate an improved technique for generating tight-binding models, in which the hopping parameters are approximately orthogonal.

\begin{figure}[h!]
\centering 
\includegraphics[width=0.5\textwidth]{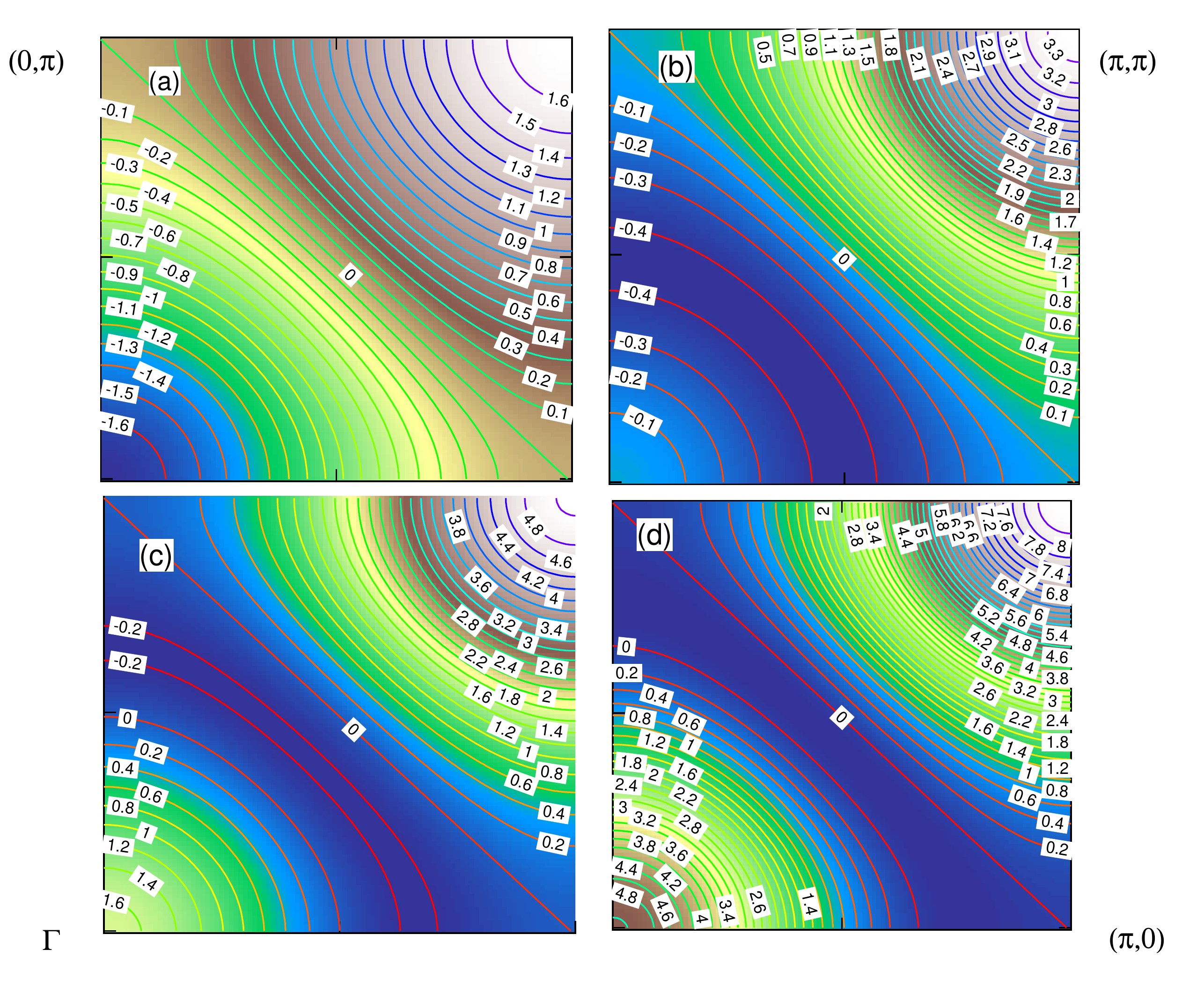}
\caption{{\bf Dispersion maps $E-E_X$ showing evolution of constant energy surfaces with $t'$ for $t''/t'=0.5.$}  $t'/t = $ (a) 0, (b) -0.5, (c) -1, and (d) -0.2.  Constant energy contours are labeled in eV.}
\label{fig:A8}
\end{figure}

Here, we generalize the analysis of Ref.~\onlinecite{VHS2} for finite $t''$.  The susceptibility divergence arises from the energy integral in Eq.~\ref{eq:1}, with $\epsilon_k$ near, e.g., $(\pi,0)$, while $\epsilon_{k+Q}$ is near $(0,\pi)$.  For the $t-t'-t''$ model, the $t'$ and $t''$ terms cancel, and the denominator becomes $\epsilon_k-\epsilon_{k+Q}=-4t(c_x+c_y)$.  Near $(0,\pi)$, $c_x+c_y\sim k_x^2-k_y^{'2} \sim k^2cos(2\phi)$, with $k_y'=\pi-k_y$.  Then the integral $\int kdk/k^2$ produces a logarithmic divergence, which can be cut off by, e.g., the temperature $T$. The angle integral $\int d\phi/cos(2\phi)$ also can diverge, if $\phi\rightarrow\pi/4$, leading to a second $ln(T)$ factor.   However, the fermi functions in Eq.~\ref{eq:1} can cut off this divergence, since at low $T$ they only allow scattering from full to empty states, and hence cut the $\phi$ integral off at the Fermi surface.  To test this possibility, we rewrite Eq.~1 by using $cos(2\theta)=2cos^2(\theta)-1$, as
\begin{equation}
\epsilon_k=-2t(c_x+c_y)-(t'+2t'')(c_x+c_y)^2+(t'-2t'')(c_x-c_y)^2
+4t''.
\label{eq:4}
\end{equation}
Thus, when $t''\ne 0$, the cutoff of Eq.~\ref{eq:4} becomes $k_BT_{eh}=|t'-2t''|$, and when $2t''=t'$, the susceptibility has a $ln^2(T)$-divergence independent of $t'+t''$, leading to a new reference family of anomalous VHSs, \Fref{fig:A8}.

Figure~\ref{fig:A8} shows that the dispersion changes strongly with $t'$, whereas the contours of constant energy are independent of $t'$, since they represent $c_x+c_y=R=constant$.  Consequently, for all values of $t'$, the zone diagonal $(0,\pi)\rightarrow (\pi,0)$ lies at the VHS, $E_F=E_X$, ensuring a square Fermi surface and a $ln^2(T)$-divergence of the $(\pi,\pi)$ susceptibility.  However, there is still a competition of $(\pi,\pi)$ and $\Gamma$ susceptibilities, except now in terms of high-order VHSs.  Note that for $t'=0$ (the conventional Hubbard model) the dispersion is electron-hole symmetric.  For $t'<0$, the dispersion becomes asymmetric, with a local peak developing at $\Gamma$, leading to a Mexican hat potential, which moves toward the $(0,\pi)\rightarrow (\pi,0)$ line as $|t'|$ increases.  The minimum of the Mexican hat produces a flat band quite similar to that found in secondary high-order VHSs.  

This similarity is not accidental.  For a mean-field model of $Q=(\pi,\pi)$ AFM order, the dispersion in the AFM phase becomes $E_{\pm}= \epsilon_+\pm\sqrt{\epsilon_-^2+U^2/4}$, where $\epsilon_{\pm}=(\epsilon_k\pm\epsilon_{k+Q})/2$.  When $U$ is large, the square-root term becomes $U/2 +\epsilon_-^2/U$.  In the $t-t'-t''$ families, $\epsilon_-=-2t(c_x+c_y)$, and the square-root term becomes $U/2 +J(c_x+c_y)^2$, $J=4t^2/ U$, having the same nonlinear term as Eq.~\ref{eq:4}.

Thus, remarkably, whereas the DOS is highly sensitive to the Fermi surface, the $q=(\pi,\pi)$ susceptibility obeys the simpler result of Eq.~\ref{eq:4}, and can only have $ln$ or $ln^2$ divergences in the $(\pi,\pi)$ susceptibility.  We note that the original Hubbard model required extreme fine tuning (all hopping parameters except $t$ equal to zero), calling into question the relevance of a square Fermi surface and resultant $ln^2$ divergence.  By showing that a square Fermi surface can be found along an entire cut in parameter space, we have greatly enhanced the likelihood that such a feature, with accompanying high-order VHSs, might be experimentally observed.  

Finally, we have defined our parameter space in terms of just the three nearest-neighbor hopping parameters, and for this to be of value, it is necessary that higher-order hopping parameters have negligible effects in most cases.  This new, `Hubbard' family offers a simple test: can a line of square-Fermi-surface states persist as higher order hopping parameters are included?  We explore this issue by looking at the most general TB model with terms of order $c_i^3$ ($E_3$) and $c_i^4$ ($E_4$), $\epsilon_{k,4} = \epsilon_k +E_3 +E_4$.  Here $\epsilon_k$ is given by Eq~1 and
\begin{multline}
E_3=-2t_3(c_{3x}+c_{3y})-4t_4(c_{2x}c_y+c_{2y}c_x)\\
=2a_3(c_x+c_y)-4A_3(c_x+c_y)^3-4B_3(c_x-c_y)^3,
\label{eq:4a}
\end{multline}
with $a_3=t_3+6t_4$, $A_3,B_3=t_3\pm t_4/3$,
\begin{multline}E_4=-2t_5(c_{4x}+c_{4y})-4t_6(c_{3x}c_y+c_{3y}c_x)-4t_7c_{2x}c_{2y}\\
=e_{40}-a_4(c_x+c_y)^2-b_4(c_x-c_y)^2-2A_4(c_x+c_y)^4\\
-2B_4(c_x-c_y)^4
-4C_4(c_xc_y)^2,
\label{eq:4b}
\end{multline}
with $e_{40}=-4(t_5+t_7)$, $a_4,b_4=4(t_5+t_7)\mp 3t_6$, $A_4,B_4=4t_5\pm t_6$, $C_4=4(t_7-6t_5)$.  Finally, since $c_xc_y=[(c_x+c_y)^2-(c_x-c_y)^2]/4$, $\epsilon_{k,4}$ can be written as a polynomial in $(c_x+c_y)$ and $(c_x-c_y)$.  Eliminating all the terms in $(c_x-c_y)$ leaves behind a polynomial in $(c_x+c_y)$, a 4th order generalization of the Hubbard reference family.

\subsection{Double Taylor series expansion}
We note in passing that our new series expansion, a Taylor series in the two variables $X=c_x+c_y$ and $Y=c_x-c_y$, should be a significant improvement over the conventional neighbor-by-neighbor tight-binding model.  We see from Eqs.~\ref{eq:4a} and~\ref{eq:4b} that each additional neighbor introduces two kinds of correction: new terms in $X$ and $Y$ (terms whose coefficients are capital letters on right-hand side of equations) and renormalizations of lower-order terms (terms with lower-case coefficients).  In contrast, the new series is essentially a symmetry-corrected Fourier series, ensuring that the new terms added at each order are orthogonal to the lower-order terms, and will be accepted only if they genuinely improve the fit to the dispersion.  Moreover, the series in X is electron-hole symmetric while the series in Y is not.  Only the latter contributes to shifting the VHS from half-filling.

For the monolayer cuprates, Eq.~\ref{eq:4} becomes
\begin{equation}
\epsilon_k=-2t(c_x+c_y)+2t'[(c_x-c_y)^2-1].
\label{eq:4b}
\end{equation}
For bilayer cuprates, hopping between the two layers making up the bilayer can lead to significant modifications of Eq.~\ref{eq:4b}, while inter-bilayer hopping remains weak.  In this case, the resulting dispersions depend on the layer stacking.  For YBCO, the CuO$_2$ layers stack Cu above Cu, and the bilayer hopping leads to a constant splitting of the bonding and antibonding bands, with no change of $t'$.  In contrast, many cuprates, including Bi2212, have a body-centered tetragonal stacking, leading to an approximate dispersive interlayer hopping\cite{PWA}
\begin{equation}
\epsilon_{bi}=\pm 2t_{bi}(c_x-c_y)^2.
\label{eq:4c}
\end{equation}
Combining with Eq.~\ref{eq:4b}, this leads to a splitting of the bonding and antibonding bands of $\pm 2t_{bi}$, and a change of dispersion, $t'\rightarrow t'\pm t_{bi}$.  More detailed analyses for particular cuprates are found in Ref.~\onlinecite{RSM1b}.

For Bi2212, this shifts the antibonding VHS into a parameter range more suitable for high $T_c$, while shifting the bonding band off of the $(\pi,\pi)$ plateau, leading to its distinctive square-cross fermi surface, as in Fig.~\ref{fig:6}(a).